\documentclass[11pt]{article}
\usepackage{graphicx}
\usepackage{amsmath}
\usepackage{wrapfig}

\input{tcilatex}

\begin{document}

\author{\textbf{Scott M. Hitchcock} \\
National Superconducting Cyclotron Laboratory (NSCL)\\
Michigan State University, East Lansing, MI 48824-1321\\
E-mail: hitchcock@nscl.msu.edu}
\title{'\textbf{Photosynthetic' Quantum Computers?}}
\date{August 19, 2001}
\maketitle

\begin{abstract}
Do quantum computers already exist in Nature? It is proposed that they do.
Photosynthesis is one example in which a 'quantum computer' component may
play a role in the 'classical' world of complex biological systems. A
'translation' of the standard metabolic description of the 'front-end' light
harvesting complex in photosynthesis into the language of quantum computers
is presented. Biological systems represent an untapped resource for thinking
about the design and operation of hybrid quantum-classical computers and
expanding our current conceptions of what defines a 'quantum computer' in
Nature.
\end{abstract}

\section{Introduction}

It was recently expressed at the \textbf{Quantum Applications Symposium (Ann
Arbor, MI, July 1-3, 2001) }during a panel discussion that there were no
examples of 'quantum computers' in nature. This author would like to point
out that there are quantum computers everywhere and our lives depends on
them.

We begin with a brief description of what we mean by 'quantum logic gates'
and 'quantum computers' in general. Any system in which a 'quantum' signal
such as a photon, exciton, phonon, fundamental particle, atom and even
molecule (in special cases) is 'input' or 'detected' and is then 'processed'
or 'computed' into one or more 'output' or 'emitted' quantum or 'classical
'signals is a quantum logic gate. If the input signal is held for later use,
then the 'gate' can be a quantum information 'register', 'accumulator' or
'memory'. The combination of logic operations \emph{on} an \textbf{%
info-state }created by the detected 'signal' and its storage in memory can
be properties of a single physical 'gate'. New info-states (and therefor
'information') can be created by logic (physical) operations on other
info-states in these 'gates'. An 'info-state' in this paper is the general 
\emph{state }of any system with any kind of 'information content'. The
content of the information is a physical property of the system such as the
energy or geometric configuration. The content of a gate is the
configuration information for the excited state capable of creating a signal
upon reconfiguration.

This gate may be a set of quantum logic gates acting as a single quantum
logic gate which we call a Collective Excitation Network or CEN \cite
{hitchcock2}. A quantum computer is any system which can process 'quantum'
(or 'classical') signals and quantized information states into one or more
quantum signals or classical states. The flow of quantum information through
a given sequential pathway of gates and connections in a causal network
creates a Sequential Excitation Network or SEN \cite{hitchcock3}. Causal
networks are SENs in which 'output' information is created by the
reconfiguration processes of unstable or 'triggered' states of 'gates'.
These nodes in the network are the local '\emph{cause}' (source or Feynman
Clock \cite{hitchcock}) components. The signals propagate in the vacuum or
along restricted permanent or temporary physical pathways ('circuits') to
detectors that convert the signals info-state into a local 'effect' (excited
state of the Feynman Clock detector or quantum logic gate) in a sequential
'history'.

The 'time' associated with the quantum information flow in quantum computers
is created by 'time labeling' states of gates by pairing them with standard
clock 'quantum' signals in a process called 'signal mapping' \cite
{hitchcock4} by a time or \textbf{'T'-computer }\cite{hitchcock5}\textbf{. }%
The T-Computer can be a 'front-end' quantum computer that can be controlled
by an observers 'classical' computer. It can be used to time label
'cause-effect' relationships between qubits, qwords (coupled sets of qubits)
and classical bits and words in shift and memory registers. The details of
quantum 'clocks' for quantum computers are being investigated by the author.

Examples of natural 'quantum computers' include the retinal photodetectors
(vision) in animals \cite{eye} and the light harvesting or 'antennae'
apparatuses in plants and some bacteria \cite{cells}, \cite{simple}, \cite
{ritz}, \cite{ritz2}, \cite{tpp}. Components of these detection systems act
as quantum or Feynman Clock logic 'gates' detecting incoming 'signals',
processing them, and creating output signals that propagate energy and
information in metabolic causal networks \cite{hitchcock}, \cite{hitchcock2}%
, \cite{hitchcock3}. We will focus on the 'quantum computer' aspects of a
highly simplified model of 'Phase I reactions' \cite{cells}\ in the
membranes of photosynthetic units or PSUs in plants as an example. PSUs are
'causal networks' acting as 'quantum' and chemical energy 'ladder' computers
processing photons into output signals such as excitons, electrons, ions and
'energy' transfer molecules like ATP.

The 'computation' of photons into bioavailable energy begins with the
conversion of photons into 'exciton' resonances in the light harvesting
antennae. These exciton signals are then sent along the 'wiring' of the PSU
as electron 'currents' between molecular structures. The continued energy
flow in the PSU represents a flow of information that is transformed into
various kinds of 'signals' such as ions and molecules in excited states by
the 'logic' of the biochemical structure of the PSU computer. The question
of whether or not energy transfers in the later stages of photosynthesis is
a 'quantum' or 'classical' process depends on the signal type. Chemical
signals generally fall into a 'classical' description if quantum mechanics
is not necessary to characterize the physical state of the signal and the
system with which it is interacting. The processing of 'quantum information'
by living systems is one of the foundations for the field of quantum biology.

At the computational nodes in the photosynthetic causal network where
chemical or physical transport of energy can be described without quantum
mechanics, we see a classical computational model may be more effective.
Careful examination of the quantum to classical transition at the mesoscopic
scale represent opportunities for designing novel hybrid biological and
physical/chemical interfaces between quantum and classical computers. In a
complex system like the purple bacteria, we see a mix of both quantum and
classical computer elements depending on the mechanism of energy
(information) transfer. The complex interactions of 'signals' with 'noise'
in these hybrid biological quantum-classical information processing systems
raises issues of metabolic efficiency of both initial signal detection and
subsequent 'logic' operations performed on the information flowing through
causal networks in complex systems \cite{me1}, \cite{me2}.

The propagation of information via different forms of signals in a causal
network tells us that signals and their information content can take various
'quantum' as well as 'classical' forms. This opens the possibilities of
looking at quantum computing systems in the broader context of biological
systems. The detection of single photons by the eye \cite{eye} entails
'computation' of both spectral and geometric (spatial and motion)
information from the 'observed' system. The photon detectors in the retina
act as 'front end' quantum computers linked to the 'classical' neural
network of the optic nerve and visual cortex in the brain. Front end
'quantum detectors' and 'computation' systems are one of natures fundamental
solutions to the problem of survival by providing living systems with a high
sensitivity to environmental data.

\section{Photosynthesis and the 'Computation' of Photons into Bio-Energy}

There are many ways to transfer energy and therefore 'information' as
signals originating in unstable or excited quantum systems (Feynman Clocks)
such as atoms and molecules. These include Excitons, Luminescence, Physical
Quenching, Ionization, Dissociation, Direct Reaction Charge Transfer,
Isomerization, Inter-molecular energy transfer, Intra-molecular
Radiationless transition energy transfer, and Phosphorescence \cite{pchem}.
In the case of photosynthesis, the 'final' signal we will examine is the ATP
molecule. ATP represents the energy 'source' for the production of glucose
and oxygen which are essential for '\emph{chemotrophs}' like us.

The light harvesting molecular antennae in the PSU of 'phototrophs' (algae,
plants and photosynthetic bacteria) detect photons and convert them directly
into usable energy for the organism via the synthesis of ATP. ATP fuels the
metabolism of phototrophic cells which includes the synthesis of large
amounts of sugars like glucose. The glucose can be further processed by
polymerization into storage macromolecules such as starch and cellulose used
in cell walls. These molecules can in turn be used as energy sources for ATP
in photosynthetic cells of most bacterial species, protozoa, fungi, and
animal cells by 'digestion'. We can see that we are intimately connected to
the quantum world of photons through the myriad of complex causal networks
we call 'life'.

\subsection{The 'Quantum Computer' Component of PSUs}

Photosynthesis is part of the 'energy processing' causal network essential
to all life \cite{tpp}. The conversion of starlight into usable biological
energy begins with photosynthesis in '\emph{phototrophs}' \cite{cells}. The
simplified model used here does not account for real details such as the
cyclic behavior of complex chemical pathways controlled by feedback
mechanisms.

Photon absorption triggers atomic and molecular reconfigurations. These
changes can be expressed as different types of physical and chemical
processes. Some of the energy conversions occur by dissociation ionization,
direct reactions, charge transfer, excitons, isomerization, intermolecular
energy transfer, intramolecular or radiationless transfer, luminescence,
fluorescence, phosphorescence, diffusion, physical quenching, resonant chain
reactions (SENs), internal conversion (e.g. decoherence) and 'thermal'
effects in unstable systems \cite{pchem}.

We will now briefly outline a \emph{simplified} version of the
photosynthetic quantum computer (PQC) components of the 'purple bacteria' as
an example. A 'translation' of the standard metabolic descriptions of the
photosynthetic apparatus in the intracytoplasmic membrane of purple bacteria 
\cite{simple} into quantum computer notation\cite{eiqc},\cite{phys229}, \cite
{qcqi}, is given below.

\textbf{Excitation transfer in the PSU of the Purple Bacteria} starting with
photon detection resulting in exciton signals propagating along the quantum
computer network to the RC 'quantum/classical' processing 'gate'.

The photon and exciton signals can be described with quantum mechanics
applied to electronic excitation in the PSU energy ladder. The following
equations are only illustrative of the \textbf{intercomplex} process of
energy and information transfer \emph{between} the complexes and the exact
details of the \textbf{intracomplex }'computation' \textbf{\ }processes 
\emph{inside} the membrane '\emph{gates}' of the PSU computer are beyond the
scope of this paper. The reader can find excellent information at the
following web pages: http://www.ks.uiuc.edu/ and
http://www.life.uiuc.edu/govindjee/paper/gov.html

The overall 'classical' \emph{chemical} equation of photosynthesis for
plants is \cite{cells}:

\begin{equation}
6H_{2}O+6CO_{2}+(light)\Rightarrow C_{6}H_{12}O_{6}+6O_{2}
\end{equation}

We will focus on the reactions in the well studied purple bacteria (e.g.
Rhodobacter (Rb.) sphaeroides) to illustrate the '\emph{front end}' quantum
computer aspects of the generally 'classical' energy transfer processes in
the remainder of the photosynthetic causal network leading to the conversion
of ADP into ATP.

The process begins with the detection of photons which are resonantly
absorbed by light harvesting complexes in the membrane of the PSU. For
purple bacteria, these can be any one or a combination of primary
resonances. The PSII complex can detect photons at 800 and 850 $\unit{nm}$
by \textbf{bacteriochlorophylls} (\textbf{BChls}). It can also detect 500 $%
\unit{nm}$ photons with the lycopene or carotenoid backbone supporting the
B800 and B850 BChl detector sites. The intracomplex carotenoid detection of
photons is too complex to describe here and will only be mentioned. For more
details, see \cite{ampb}.

We will look at a \emph{simplified} metabolic pathway involving the
detection of an 800 or an 850 $\unit{nm}$ photon. We will also ignore the
photon detector 'accumulator' or 'register' function of the complex in
building an \textbf{exciton} resonance by the complex with many (8 or more?)
separate detections. The \emph{intercomplex} transmission of energy is from
one complex to another interior \textbf{LHII }or \textbf{LHI} complex on its
way to the \textbf{reaction center }(\textbf{RC}). The LHI complex can also
detect photons at 875 $\unit{nm}$.

The \textbf{RC} represents the site or 'gate' where \emph{quantum information%
} is converted into \emph{classical chemical information}. The purpose of
this paper is to suggest how we may think about the biological processing of
photons into life energy as having 'quantum' as well as 'classical' computer
aspects. The subtle details of 'quantum biology' demand a much more careful
treatment that can be given here. The reader is asked to see the references
for such a treatment.

\subsubsection{A Simple Example of 'Front-end' Quantum Computer 'Logic' in
the PSU of Purple Bacteria}

\ We will look at the maximum information path in the PSU from the detection
of an 800 or an 850 $\unit{nm}$ photon by an outer LHII. The energy/signal
will be transferred to an inner LHII by an 'exciton'. This inner LHII will
then transfer the exciton to the LHI. The LHI will then transfer the
energy/signal to the reaction center, RC, where it is converted into
'available ' energy as electron-hole separation for the 'quantum' processing
of $\mathbf{H}^{+\text{ }}$ions (protons) and electrons with a 'classical'
binding of cytoplasmic \textbf{quinone, }$\mathbf{Q}_{B}$\textbf{\ , }to the 
\textbf{RC}. The $\mathbf{Q}_{B}$ is reduced to hydroquinone, $\mathbf{Q}_{B}%
\mathbf{H}_{2}$, and then released.

The $\mathbf{Q}_{B}\mathbf{H}_{2}$ is then processed by the $\mathbf{bc}_{1}$%
complex in an exothermic reaction that pumps protons across the membrane
into the periplasm. During this operation, electrons are transmitted to the $%
\mathbf{RC}$ from the $\mathbf{c}_{2}$ cytochrome complex on the periplasmic
side to the $\mathbf{bc}_{1}$ to the $\mathbf{c}_{2}$ 'gate' on the
periplasmic side to the $\mathbf{RC}$.

The electron transfer across the membrane induces a proton gradient that
drives the 'classical' synthesis of $\mathbf{ATP}$ from $\mathbf{ADP}$ at
the $\mathbf{ATPase}$ 'gate'. This is the point at which energy for other
complex 'life' processes becomes 'classically' or chemically available.

The detection of the photon, $\left| \lambda _{II}^{(2)}\right\rangle $, by
an outer LHII complex resulting in an excited state, $\left| LHII_{\ast
}^{(2)}\right\rangle \left| LHII_{\ast }^{(2)}\right\rangle $, is
represented by:

\begin{equation}
\left| \lambda _{II}^{(2)}\right\rangle +\left| LHII_{0}^{(2)}\right\rangle
\Longrightarrow \left| \lambda _{II}^{(2)}\right\rangle \otimes \left|
LHII_{0}^{(2)}\right\rangle =\left| LHII_{\ast }^{(2)}\right\rangle 
\end{equation}

After the needed number of photons are detected, the excited state of this
LHII Complex represents the \emph{collective state} of an \textbf{%
intracomplex} network of the individual Feynman Clocks (B800, B850, and
lycopene FCs) which 'decays' in a 'finite lifetime' (computed using a
T-computer time labeling process or the like) with the creation of an \emph{%
exciton signal}.

Next we have the 'decay' of this info-state and the creation of an exciton, $%
\left| \epsilon _{II\rightarrow II}\right\rangle $, which transfers
information to the inner LHII complex:

\begin{align}
\left| LHII_{\ast}^{(2)}\right\rangle & \Longrightarrow\left|
LHII_{0}^{(2)}\right\rangle \otimes\left| \epsilon_{II\rightarrow
II}\right\rangle \Longrightarrow \\
\left| LHII_{0}^{(2)}\right\rangle +\left| \epsilon_{II\rightarrow
I}I\right\rangle & \Longrightarrow\left| LHII_{0}^{(1)}\right\rangle
\otimes\left| \epsilon_{II\rightarrow II}\right\rangle \Longrightarrow\left|
LHII_{\ast}^{(1)}\right\rangle
\end{align}

The excited state of the inner LHII decays or transfers the energy in a
similar manner to the LHI. We have:

\begin{align}
\left| LHII_{\ast}^{(1)}\right\rangle & \Longrightarrow\left|
LHII_{0}^{(1)}\right\rangle \otimes\left| \epsilon_{II\rightarrow
I}\right\rangle \Longrightarrow\left| LHII_{0}^{(1)}\right\rangle +\left|
\epsilon_{II\rightarrow I}\right\rangle \\
& \Longrightarrow\left| \epsilon_{II\rightarrow I}\right\rangle
\otimes\left| LHI_{0}\right\rangle \Longrightarrow\left| LHI_{\ast
}\right\rangle
\end{align}

The inner LHI excited info-state provides the information for the creation
of an exciton signal that propagates inward to the RC:

\begin{align}
\left| LHI_{\ast}\right\rangle & \Longrightarrow\left| LHI_{0}\right\rangle
\otimes\left| \epsilon_{I\rightarrow RC}\right\rangle \Longrightarrow\left|
LHI_{0}\right\rangle +\left| \epsilon_{I\rightarrow RC}\right\rangle \\
& \Longrightarrow\left| \epsilon_{I\rightarrow RC}\right\rangle
\otimes\left| RC_{0}\right\rangle \Longrightarrow\left| RC_{\ast
}\right\rangle
\end{align}

Now we see how the original photon signal information has propagated
alternating between 'free' signals and info-states of complexes delivering
energy and information to the RC. In summary, the photon has been processed
into an unstable info-state of the outer LHII, then into an inwardly
propagating \emph{exciton signal} down the 'energy ladder'. This is followed
the excitons detection by an inner LHII which is creates an unstable
info-state of that LHII complex. This info-state is processed into another
(in the case of direct transfer, the first and second excitons between LHIIs
can be considered the 'same') exciton signal which propagates to the LHI
complex. This creates an unstable info-state of the LHI which in turn decays
into a ground state by the transfer of information via an exciton to the
inner RC core of the LHI. The 'excited' state of the RC is where the
'output' signal of this front-end quantum computer meets the 'classical'
chemical computer components that further process the original photons
information into the chemistry of 'life'.

We can see that the distinction between quantum photon and exciton processes
and classical 'chemical' processes becomes 'fuzzy' in the $\mathbf{RC}$. The
transport of electrons and protons can be treated classically when their
gate to gate path-lengths are much larger than the size of the systems in
which they are 'processed' or 'generated'. The 'chemical' processes reflect
the fact that the transfer of energy and information is mediated by 'mobile'
systems whose physical scale with respect to the quantum information content
allows it to be described with classical physics. This can be thought of as
classical if the magnitude of the 'action', $\mathbf{A=}$ $\left| \mathbf{%
\vec{p}\cdot \vec{r}}\right| \mathbf{=E\cdot t}$ involved with the 'signal'
and 'detector' interaction is $\mathbf{A\gg }1.05457266\times 10^{-34}\unit{J%
}\unit{s}=\hbar $ \cite{white}.

\subsubsection{'Time' and Info-flow in Quantum Computers}

The \emph{state information} is transferred from gate to gate as an '\textbf{%
info-state}'. The creation of a new info-state in a gate can be 'time
labeled' using the signals from a standard clock by a method of 'signal
mapping' with a \textbf{T-computer} \cite{hitchcock5}. Clock signals can
also be used to 'drive' computational processes.

The quantum arrows of time or QATs are 'constructions' built from
information originating in the Feynman Clock (FC), Collective Excitation
Network (CEN) and quantum Sequential Excitation Network (SEN) \cite
{hitchcock3} processes in a cell and 'computed' into 'time labels' by the
observing system or a higher order computing system of which the quantum
computer is a part. The quantum computer components of the PSU represent the
building blocks of larger biological networks such as trees with their
energy and chemical fluid transport systems creating daily and seasonal
'arrows of time' for overall complex system configurations of a
superposition of various metabolic causal networks. This model can be
generalized to the information processing quantum and classical computers in
animals. The control and transport systems are hierarchically built upward
in systematic complexity with benchmark plateaus of complexity at the
cellular and organ level to support the activities of the central nervous
system and the brain.

\subsection{The Creation of Novel Information States and the 'Conservation
of Information Law'}

Biological system are distinct from other evolving systems such as galaxies,
stars and planets in that they can process information and create novel and
complex forms of information embodied in 'hand-made' physical systems and
artifacts. The overall flow of information in the universe is driven by
stellar evolution. How is it that new forms of information can be created
from evolving physical systems? How do these new info-states of material
systems arise? Are they computed from previous info-states in conjunction
with some sort of information 'reservoir' of other info-states in order to
account for the differences between the initial and final configurations of
a system? We can look at the possibility that if there is a 'Conservation of
Information Law' \cite{hitchcock6} for the entire universe that new
info-states are computed using other local info-states (signals) and the
expanding 'vacuum' of space as a reservoir of information vastly larger (in
direct proportion to the total mass of the dark energy or missing mass) than
the 'visible' astronomical objects we observe. A general conservation of
information law will offer a vantage point to 'account' for the information
content and novelty of evolving physical systems while addressing question
of non-local information entanglement in quantum computers of biological and
condensed matter origins. It may also be useful in designing future hybrid
quantum and classical computer systems by providing a means for accounting
for all information involved in complex computational processes.

\section{Hybrid Quantum-Classical Computers?}

If we stay within the domain of the 'physics' approach to quantum
computation as an extension of the solid state approach to classical large
scale computers, then we are overlooking the opportunity to introduce the
adaptive and quasi-analog properties that biological quantum computers offer.

\section{Acknowledgments}

Special thanks to \textbf{Steven Beher} for valuable discussions about the
ideas in this paper during and after our attendance at the Quantum
Applications Symposium. Thanks to Toni H. for her valuable 'time'.

\noindent

\end{document}